\documentclass[twocolumn,amssymb, nobibnotes, aps, prb]{revtex4-2}
\usepackage{bm}
\usepackage{hyperref}
\usepackage{amsmath,amssymb,amsfonts,graphicx,float}

\newcommand{\beq}{\begin{eqnarray}}
\newcommand{\eeq}{\end{eqnarray}}
\usepackage{amsmath}
\usepackage{amssymb}
\usepackage[paper=letterpaper,margin=1in]{geometry}
\usepackage{braket}
\usepackage{upgreek}
\usepackage{slashed}
\usepackage{soul} 

\def\pauli{\sigma}

\usepackage{tikz}

\begin{document}

\title{Kitaev Chain with a Fractional Twist}
\author{Bora Basa$^1$, Gabriele La Nave$^2$, and Philip W. Phillips$^1$}

\affiliation{$^1$Department of Physics and Institute for Condensed Matter Theory,
University of Illinois
1110 W. Green Street, Urbana, IL 61801, U.S.A.}
\affiliation{$^2$Department of Mathematics, University of Illinois, Urbana, Il. 61801}

\begin{abstract}
The topological non-triviality of insulating phases of matter are by now well understood through topological K-theory where the indices of the Dirac operators are assembled into topological classes. We consider in the context of the Kitaev chain a notion of a generalized Dirac operator where the associated Clifford algebra is centrally extended. We demonstrate that the central extension is achieved via taking rational operator powers of Pauli matrices that appear in the corresponding BdG Hamiltonian. Doing so introduces a pseudo-metallic component to the topological phase diagram within which the winding number is valued in $\mathbb Q$. We find that this phase hosts a mode that remains extended in the presence of weak disorder, motivating a topological interpretation of a non-integral winding number. We remark that this is in correspondence with Ref.~\cite{singer} demonstrating that projective Dirac operators defined in the absence of spin$^\mathbb C$ structure have rational indices.   
\end{abstract}

\maketitle

\section{Introduction}
Dirac's proposal~\cite{dirac} of taking the square root of the Klein-Gordon equation has had a remarkable impact on theoretical physics. This simple manipulation doubled the number of particles and eventually led to the formalization of the quantum field theoretic concept of spin. With our modern understanding of the analytic-geometric aspects of spin, we know now that this proposal, while simple, is far from naive. In the context of topological materials, there has been a resurgence of Dirac's original intuition with the advent of square-root topological insulators and also square-root Weyl semi-metals. Both systems emerge by taking the square root of either an appropriate tight-binding model~\cite{arkinstall} which can lead to a new class of topological insulator that allows robust edge states with codimension larger than one~\cite{edge2,edge3,edge4,edge5,edge6,edge7,edge8,edge9} or by stacking such two-dimensional (2D) square-root higher-order topological insulators with interlayer couplings in a double-helix pattern~\cite{srws}.  While both might seem artificial, the former has been observed in a photonic cage~\cite{photoniccage}.  The key prediction of the square-root Weyl semimetal proposal~\cite{srws} is the presence of Fermi arcs and hinge states that connect the projection of the Weyl points.  Even the latter~\cite{srws} has had an experimental realization.

From a field theoretic perspective, a QFT with second-order field equations defined on a manifold that admits spin structure will indeed have a meaningful Fermionic `square-root' theory. In fact, if the parent theory is not very exotic, the local square-root counterpart will not be either. Since taking a square-root is a topologically-nontrivial manipulation of the space of sections of the parent theory, the emergence of new/exotic topological features in these square root condensed matter systems is well motivated.

The special importance of the square root is clear. Though, one could ask if the (schematic) generalization $(\cdot)^{1/2}\to(\cdot)^{m/n}$ can be suitably made rigorous in a way that is applicable to topological condensed matter systems. We answer this question in the affirmative by a central extension of the Clifford algebra that closes for rational powers of the Dirac matrices. This new algebra is compatible not with the ordinary Dirac operator but with a pseudo-differential analog that is locally Dirac-like but only globalizes projectively. Such projective Dirac operators defined in the absence of a spin structure are shown to have rational topological index~\cite{singer}, albeit with theoretical machinery absent in the traditional treatment of topological Hamiltonian systems.

In this work, we realize these formal ideas in a Kitaev chain~\cite{Kitaev_2001} of spinless Fermions, chosen for its simplicity as well as its ubiquity, with the BdG doubled nearest-neighbour coupling Hamiltonians carry Pauli matrices raised to a rational power. The algebraic structures we need for our generalization live entirely in particle-hole space where we show that taking fractional powers gives rise to central extensions that lead to a fundamentally altered topological phase space.  Most notably, we report the existence of a pseudo-metallic phase marked by a dense set of mid-gap modes with rational winding and bulk gap closings occurring at a half integer winding number. Interestingly, we find delocalized modes in this pseudo-metallic region that resist localization in the presence of on-site disorder only when the untwisted theory is tuned to its topological phase. We conclude with a discussion of the field theoretic formulation of projective Dirac physics that motivates the interpretation of a rational winding number as a topological index and relate it to an extended classification scheme of topological phases.    


\section{Rational powers of BdG Hamiltonians}

One of the standard models exhibiting topological superconductivity is a chain of spinless Fermions coupled by a nearest neighbour p-wave pairing. This is the so-called Kitaev chain~\cite{Kitaev_2001} with minimal Hamiltonian,
\begin{equation}
    H = -\sum_{i}\left\{(c^\dagger_{i+1}c_{i}-\frac{\Delta}{t} c^\dagger_{i+1}c^\dagger_i+h.c.)+\frac{\mu}{t} c^\dagger_ic_i \right\}.
\end{equation}
This model, being a Fermionic theory, is constrained by Fermion parity. In general $\Delta\in\mathbb C$. Time reversal symmetry is present if the pairing is chosen in $\mathbb R$ and is broken otherwise. This puts the model in the symmetry classes~\cite{az} BDI or D respectively. 

The standard way of dealing with the pairing term is to work in the doubled Bogoliubov–de Gennes (BdG) basis, 
\beq
\psi=(c_1,...,c_N,c^\dagger_1,...,c_N^\dagger),
\eeq
leading to the Hamiltonian
\begin{equation}\label{eq:kit}
    H = -\sum_{i}\frac{\mu}{t}\psi^\dagger_i \sigma_z\psi_i+\left(\psi^\dagger_{i+1}\left(\sigma_z+\frac{\Delta}{t}i\sigma_y\right)\psi_{i}+h.c.\right).
\end{equation}
With open boundary conditions, there exist localized Majorana modes on the boundaries of the chain in the topologically non-trivial phase. The familiar phase diagram of this model can be inferred from Fig.~\ref{fig:1}(a,b). 

We view the standard BdG Hamiltonian in dimension one as arising from the representation theory of $\mathfrak{spin}_{2n}$. We will focus on the case of BdG doubled spinless Fermions, where the algebraic object of interest is the $\mathfrak su(2)$ particle-hole algebra, of which the blocks of the real-space Hamiltonians furnish representations. One possible modification of standard BdG Hamiltonians describing linearized superconducting pairing is to enlarge this algebra.  

To this end, we define a general operator power by the integral
\begin{equation}
    A^\gamma= \frac{1}{ \Gamma (-\gamma)}\int_0^\infty \frac{dt}{t^{1+\gamma}}\left(e^{-tA}-\mathbb I\right),
\end{equation}
for $\gamma\in[0, 1]$, though this requires a strictly positive spectrum. Of course, when the operator in question is Hermitian, we have the more intuitive definition,
\begin{equation}\label{eq:def}
    A^\gamma = U\text{diag}(\text{spec} A)^\gamma  U^\dagger.
\end{equation}
We now imagine taking rational powers of the BdG Hamiltonian sub-blocks in Eq.~\eqref{eq:kit}, valued in $2\times 2$ Pauli matrices.

Indeed, $\sigma_i^\gamma$ is not in $su(2)$ for $\gamma\ne 1$.  Recall that for (ijk), any permutation of $\{1,2,3\}$,
\beq
[\sigma_i,\sigma_j]=2i\varepsilon^{ijk}\sigma_k.
\eeq
To close the algebra for $\sigma_i^\gamma$, we proceed as follows. We first define $z_\gamma:=(-1)^\gamma$.  Using, Eq.~\eqref{eq:def}, we obtain the relations
\begin{equation}\label{eq:sigmas}
\begin{aligned}
\sigma_k^\gamma &= \frac{1+z_\gamma}{2}\mathbb I+\frac{1-z_\gamma}{2}\sigma_k\\
\sigma_k&= -\frac{2\sigma_k^\gamma}{z_\gamma-1}+\frac{z_\gamma+1}{z_\gamma-1}\mathbb I.
\end{aligned}
\end{equation}
These relations can be used to bring the resultant commutator
\beq\label{eq:ideal}
[\sigma_i^\gamma,\sigma_j^\gamma]=\frac{i}{2}\varepsilon^{ijk}(z_\gamma-1)^2\sigma_k,
\eeq
into the form 
\begin{equation}\label{eq:comm}
[\sigma_i^\gamma,\sigma_j^\gamma]=i\varepsilon^{ijk}\sigma^\gamma_k(1-z_\gamma)+\frac{1}{2}\frac{z_\gamma+1}{z_\gamma-1}\mathbb I,
\end{equation}
where it is evident that the algebra has a central extension with $w\mathbb I$ in the center of the group generated by the $\sigma_i^\gamma$'s. This importantly gives rise to a projective representation (due to Bargmann's theorem~\cite{bargmann}), i.e. a homomorphism of $SU(2)$ into $PGL(2,\mathbb C)$. As we will see later, this is the main feature of our paper. The topological nature-- specifically the rationality of the winding number-- is a fundamental property of the projectivity of the representations involved in the construction of the Hamiltonian.  

Particle-hole-space Hamiltonians that are valued in this new algebra via
\begin{equation}
\sum_i\sigma_i^\gamma f^i(k)
\end{equation}
are normal but anti-Hermitian. Further, if one introduces the ordinary Pauli matrices in this construction along with their fractional counterparts as in Eq.~\eqref{eq:ideal}, the resulting Hamiltonian is non-Hermitian. While breaking from Hermiticity is interesting on its own right, we opt to restore it by defining a self-adjoint rational power,
\begin{equation}
 \tilde\sigma^\gamma_k = \frac{\sigma_k^\gamma}{2}+\frac{(\sigma_k^\gamma)^\dagger}{2}.
\end{equation}
The Hermitian central extension results in commutators of the form 

\begin{align}\label{eq:sigmas1}
\tilde\sigma_k^\gamma &= \frac{1-\cos\pi \gamma}{2}\tilde\sigma_k+\frac{1+\cos\pi \gamma}{2}\mathbb I\\
[\tilde\sigma_i^\gamma,\tilde\sigma_j^\gamma]&=i\varepsilon^{ijk}\tilde\sigma^\gamma_k(1-\cos\pi \gamma)+\frac{\mathbb I}{2}\frac{1+\cos\pi \gamma}{1-\cos\pi \gamma}.
\end{align}
We note that when $\gamma\rightarrow 1$, $\tilde\sigma_k^\gamma$ goes smoothly over to $\tilde\sigma_k$.

We can now define the fractionally twisted BdG Hamiltonian for the Kitaev chain:
\begin{equation}\label{eq:kit-domain}
    H = \sum_{i}\frac{\mu}{t}\psi^\dagger_i\sigma_z\psi_i+\left(\psi^\dagger_{i+1}\left(\tilde\sigma_z^{\gamma_h}+i\frac{\Delta}{t}\tilde\sigma_y^{ \gamma_p }\right)\psi_{i}+h.c.\right),
\end{equation}
where $ \gamma_{h,p} \in\mathbb Q$ is the matrix power of the hopping and pairing Hamiltonians respectively. The couplings are not included in the power to ease comparisons between the fractional and integral models. We will consider $\gamma\equiv \gamma_p$, $\gamma_h=1$ unless stated otherwise. 

The off-diagonal Hamiltonian block can be expressed in terms of $\pauli_{y,z}$ and the identity matrix using Eq.~\eqref{eq:sigmas1}. In that regard, our model with the rational power of $\sigma_y$ can be reinterpreted as the Kitaev chain with an additional nearest-neighbour coupling term that mixes the hopping and pairing strengths:
\begin{widetext}
\begin{equation}\label{newalgebra}
    H = \sum_{i}\frac{\mu}{t}\psi^\dagger_i\sigma_z\psi_i+\left(\psi^\dagger_{i+1}\left(\sigma_z+i\frac{\Delta}{t}\sigma_y+\mathbb I i\frac{\Delta}{2 t}\frac{1+\cos\pi\gamma}{1-\cos\pi\gamma}\right)\psi_{i}+h.c.\right).
\end{equation}
\end{widetext}
We opt for the former representation of the model as the latter obfuscates the underlying algebraic structure from which the model emerges. However, Eq. (\ref{newalgebra}), appears to be more experimentally transparent. 

The choice of which Hamiltonian block to raise to a rational power is somewhat arbitrary. In discussing the phase diagram, we consider also taking powers of the hopping Hamiltonian.


\section{Topological characterization of the fractionally twisted Kitaev chain}
In momentum space, we have the generic effective Dirac Hamiltonian, $H_{D} = \sum_i f_i(k) \sigma^i$, where, roughly speaking, for a 1 dimensional Brillouin zone (BZ), the degree of the map between spheres, $f_i:S^1\to \mathbb CP^1$, characterizes the integral winding number of the Hamiltonian.  More precisely, one is interested in casting the topological classification of such Dirac-like Hamiltonians\begin{footnote}{At the level of lattice Hamiltonians, we take topological non-triviality to simply mean that there exists an obstruction to finding global sections of the occupied Hilbert sub-bundle over the BZ.}\end{footnote} as the classification of Dirac operators acting on sections of Dirac bundles. This objective is formalized using (twisted) topological K-theory to develop a periodic table of topological insulators~\cite{ktheory1, freedTwistedEquivariantMatter2013}.

\subsection{Winding number}
In the case of the BDI class in dimension 1 the Dirac fibers are particle-hole spaces and, $\text{index }\slashed D\in\mathbb Z$, corresponding to a specific topological number like the $\hat A$-genus. This topological invariant can be captured by computing the winding number of the ground-state.

In fact, more generally one can consider a family of such theories where the Dirac operators $\slashed D _\gamma$, are parametrized by some manifold which, for concreteness, we take to be $S^1.$ Then, there still holds a version of the index theorem. Namely, the geometric incarnation which is the Atiyah-Singer index theorem and it comes from globalizing  representations of the Clifford algebras by means of spin$^\mathbb C$ structures (whose existence depends only on a  topological invariant, the third Stiefel-Whitney class, $w_3(M)\in H_3(M;\mathbb Z)$), of the underlying manifold $M$. What is more remarkable is that even in the case in which such spin$^\mathbb C$ structures exist only modulo projectivization, a version of the index theorem where the index is valued in twisted K-theory and is now a rational number\cite{singer} still holds.  

In our context, we take $\gamma \in S^1= \mathbb R/\mathbb Z$, by periodicity, and we observe that in the geometric realization of the model in the Brillouin zone, the effective base space of the projective Dirac operator with parameters is  $M=\mathbb C\mathbb P^1\times S^1$. The work of Melrose et al.~\cite{singer} requires the twisted K-theory be done by means of torsion classes in $H_3(M;\mathbb Z)$ (which are absent here). We therefore cannot infer that the analytic and topological indices are equal here, since we would be twisting with non-torsion classes. Instead, we probe the topological nature of the fractional analytical index numerically in the proceeding section. Further, we contend that the main property that ensures their version of the index theorem holds requires only projective representations. We further elaborate on this in Sec.~\ref{sec:QFT}.



\begin{figure*}[t]
\includegraphics[width=0.49\linewidth]{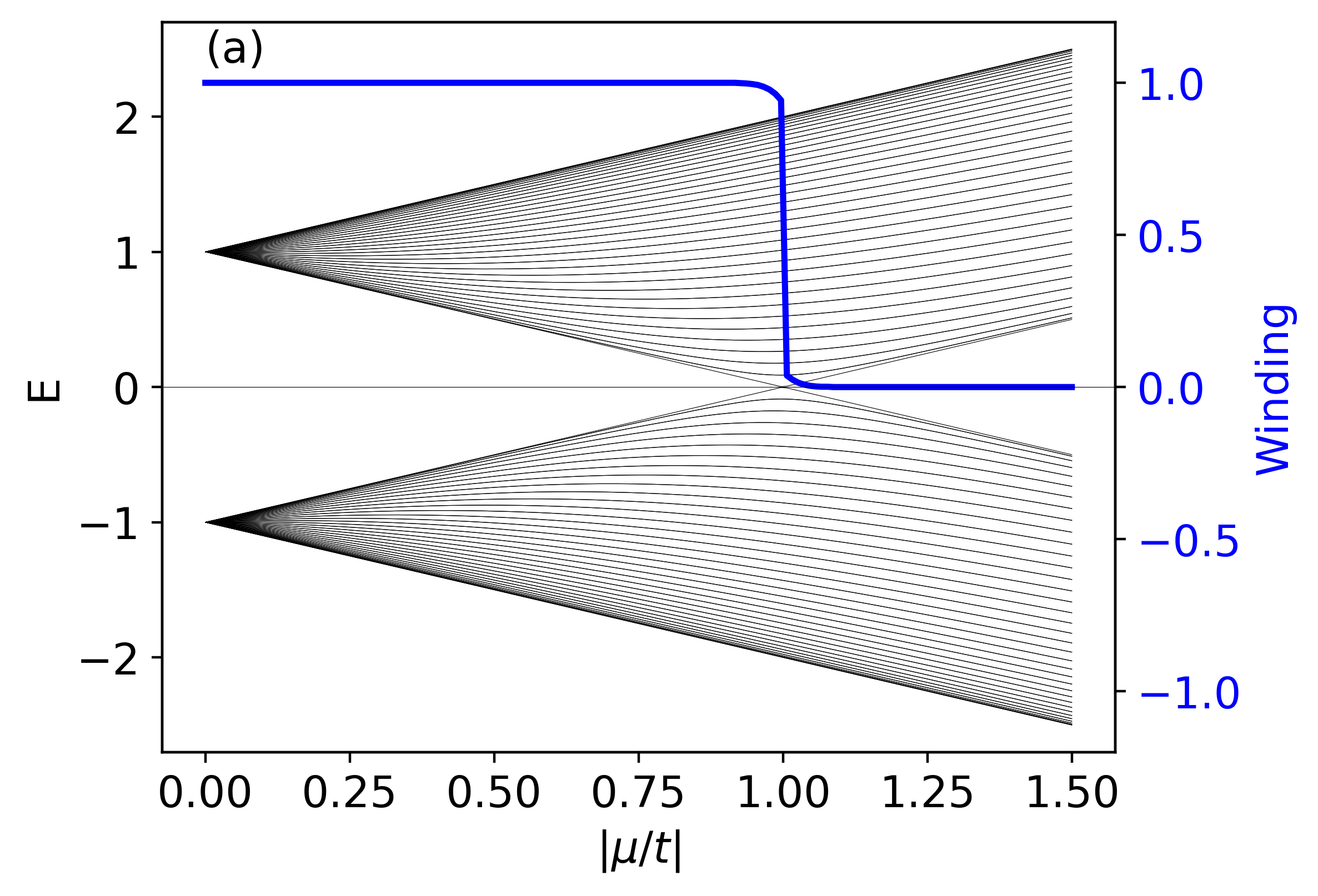}
\includegraphics[width=0.49\linewidth]{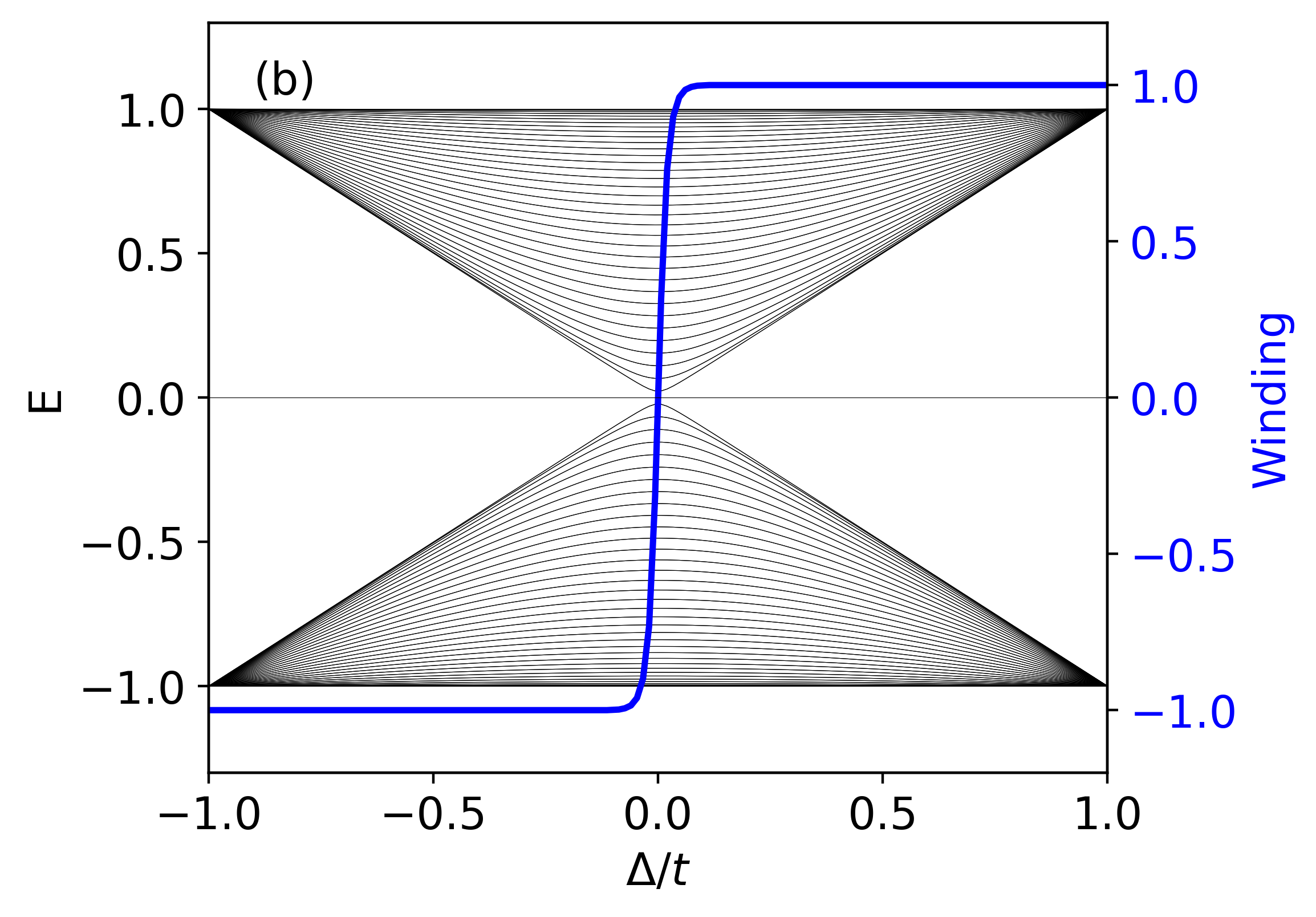} \includegraphics[width=0.49\linewidth]{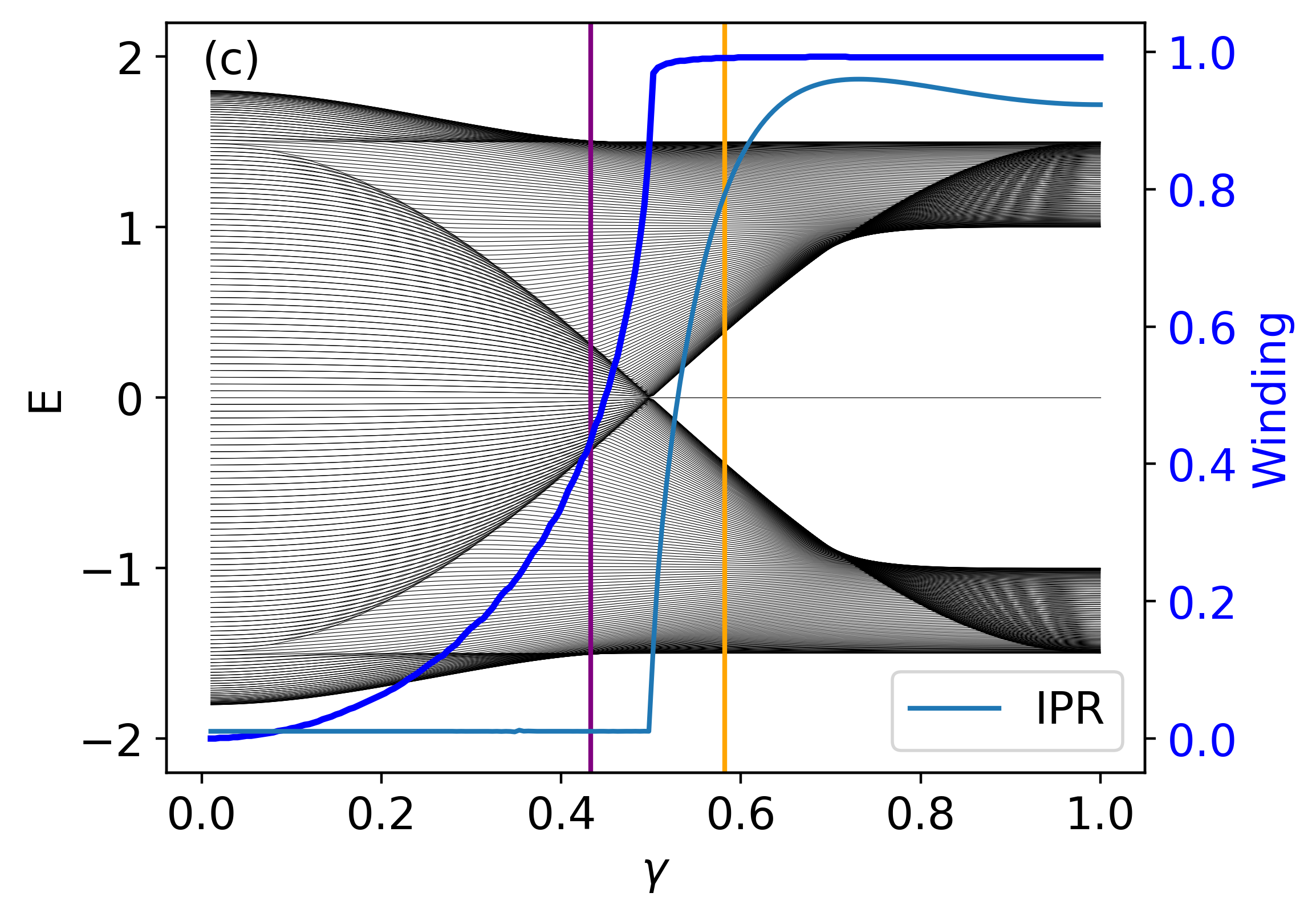}
\includegraphics[width=0.49\linewidth]{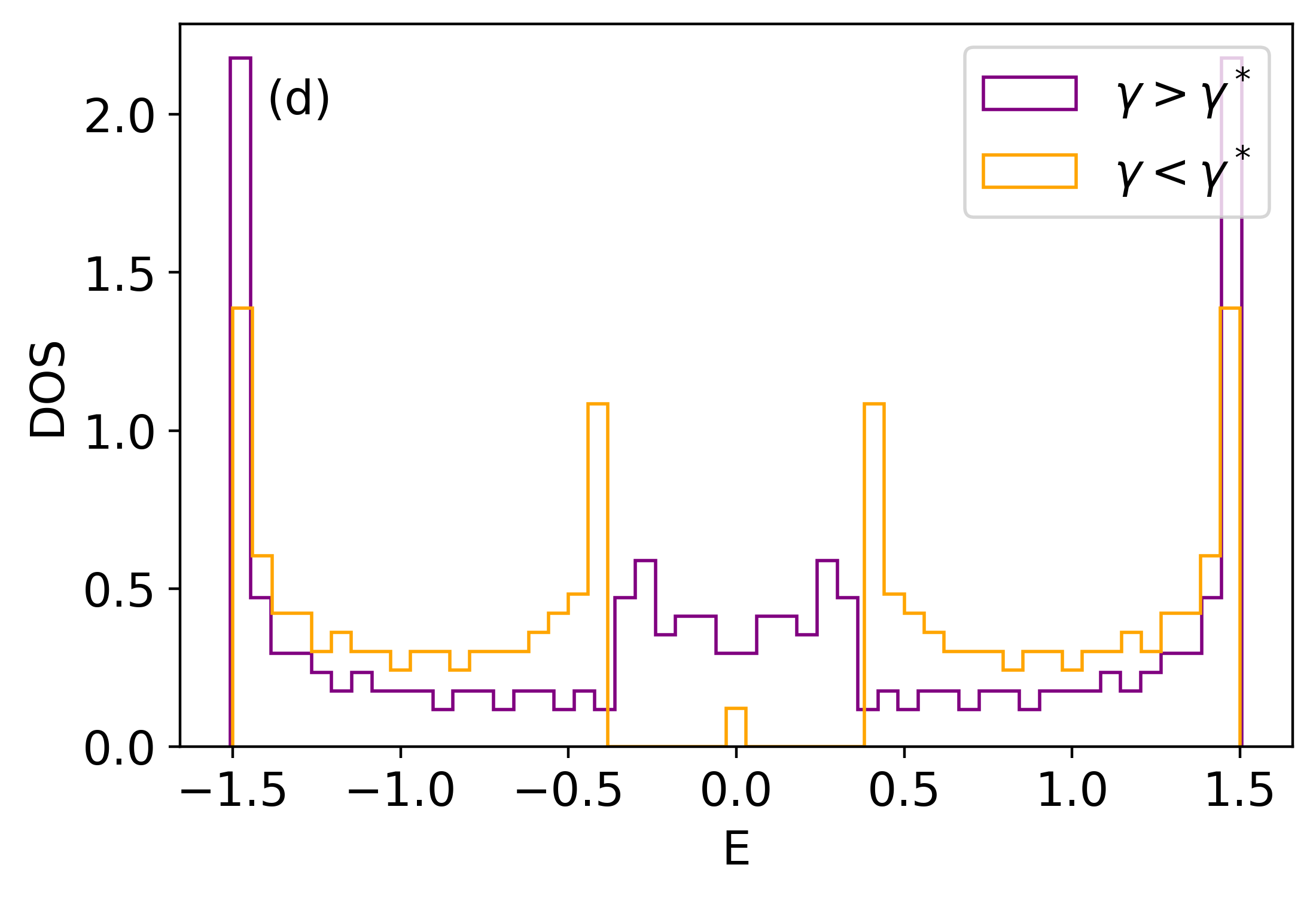}
\caption{Evolution of the spectrum of the ordinary Kitaev chain ($\gamma=1$) as a function of (a) $\mu$, for $\Delta=1$ and (b) $\Delta$, for $\mu=0$.  The standard critical point where the gap closes corresponds to $\mu=1$ at which point the winding number transitions from 1 to zero.  For any $\Delta\neq 0$, the winding number is $\text{sgn }\Delta$ and vanishes when $\Delta=0$ and the fractional Kitaev chain with (c) $\gamma$ for $\mu=0,\Delta=1$ and (d) its DOS along two energy slices. The winding number is approximated in real space based on the prescription given in~\cite{prodan2}. Beyond the appearance of the delocalized mid-gap states, the $\mu$ and $\Delta$ spectral evolution for $\gamma\in (0, 1)$ does not provide any new information. We plot also the inverse participation ratio (IPR) of the zero-mode in (c) to indicate (de)localization.}
\label{fig:1}
\end{figure*}


\subsection{Computation of the real space winding number}
Since we introduce the fractional twist in real space, we opt to compute the winding number in real space as well. To this end, we follow the algorithm developed in Ref.~\cite{prodan1,prodan2} which, by construction, can capture non-integral winding numbers (see Ref.~\cite{PhysRevResearch.3.013148} for a comparison with  other common algorthims). Following this construction, we define a homotopic flat band Hamiltonian of the form
\begin{equation}
H_0 = \begin{pmatrix}
    0 &Q_\gamma\\
    Q^\dagger_\gamma & 0
\end{pmatrix},
\end{equation}
where $Q_\gamma$ is valued in a projective representation of SU(2). Then, the winding number has the real space representation~\cite{prodan1},
\beq
\nu = \text{Tr}(Q_\gamma^{-1}[X, Q_\gamma]).
\eeq
The commutator in question is written
\beq
[X, Q_\gamma]=\sum_{l\neq 1}c_l l^XQ_\gamma l^{-X},\quad c_l\frac{l^{N+1}}{1-l}.
\eeq
Of course, these expressions hold only approximately in finite volume. 





\subsection{Phase structure}

\begin{figure*}
\includegraphics[width=0.47\linewidth]{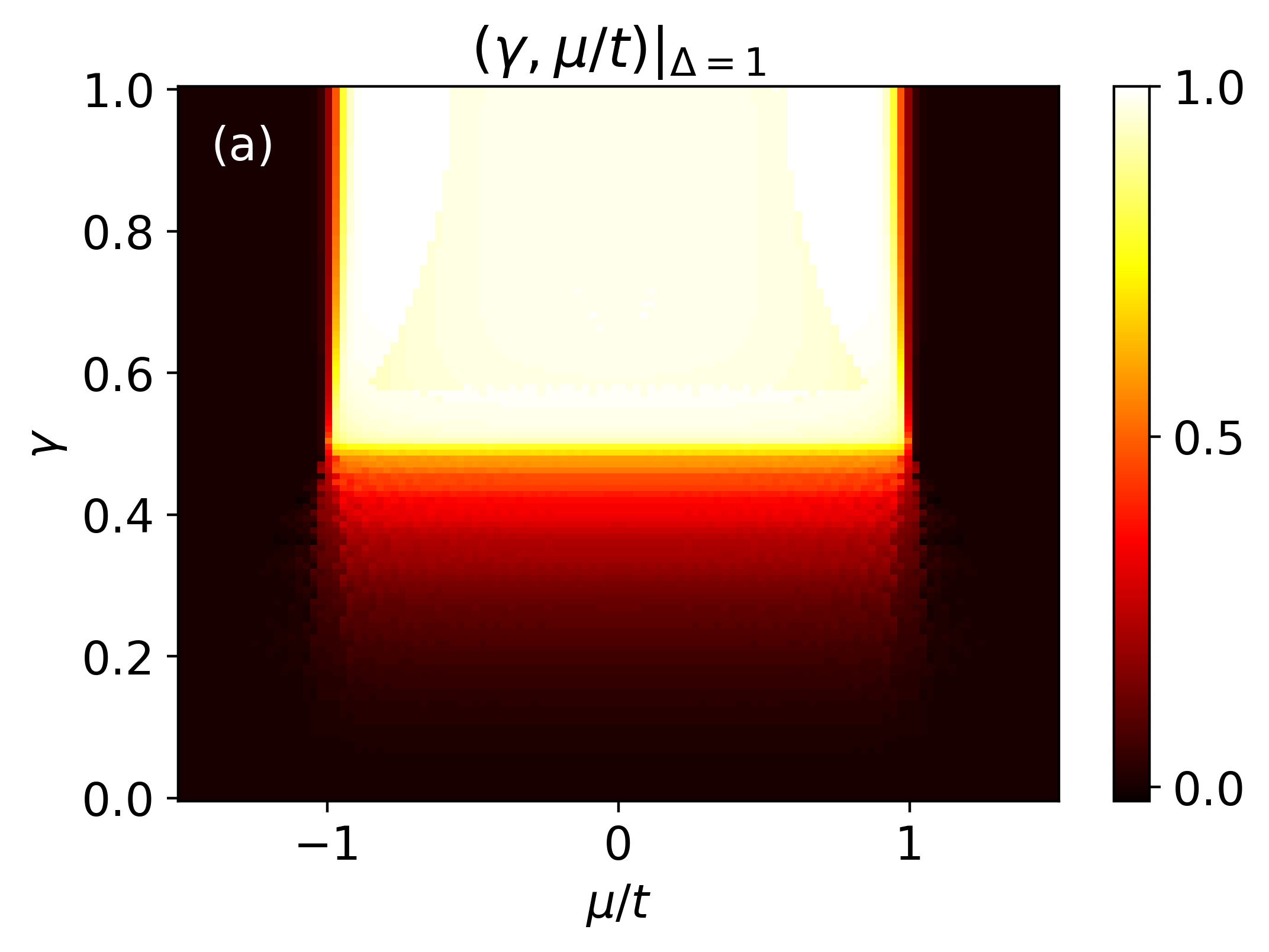}
\includegraphics[width=0.47\linewidth]{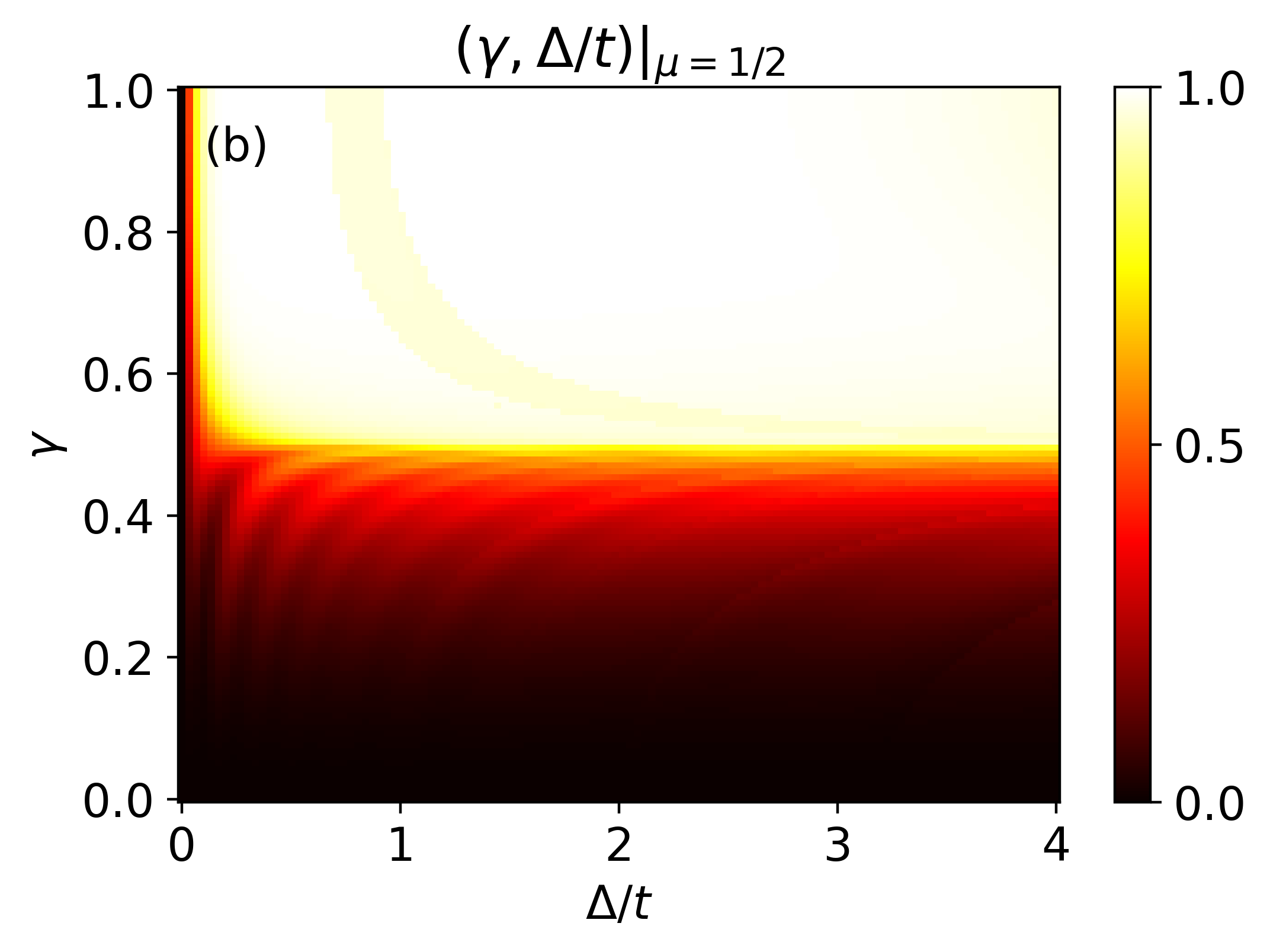}
\includegraphics[width=0.47\linewidth]{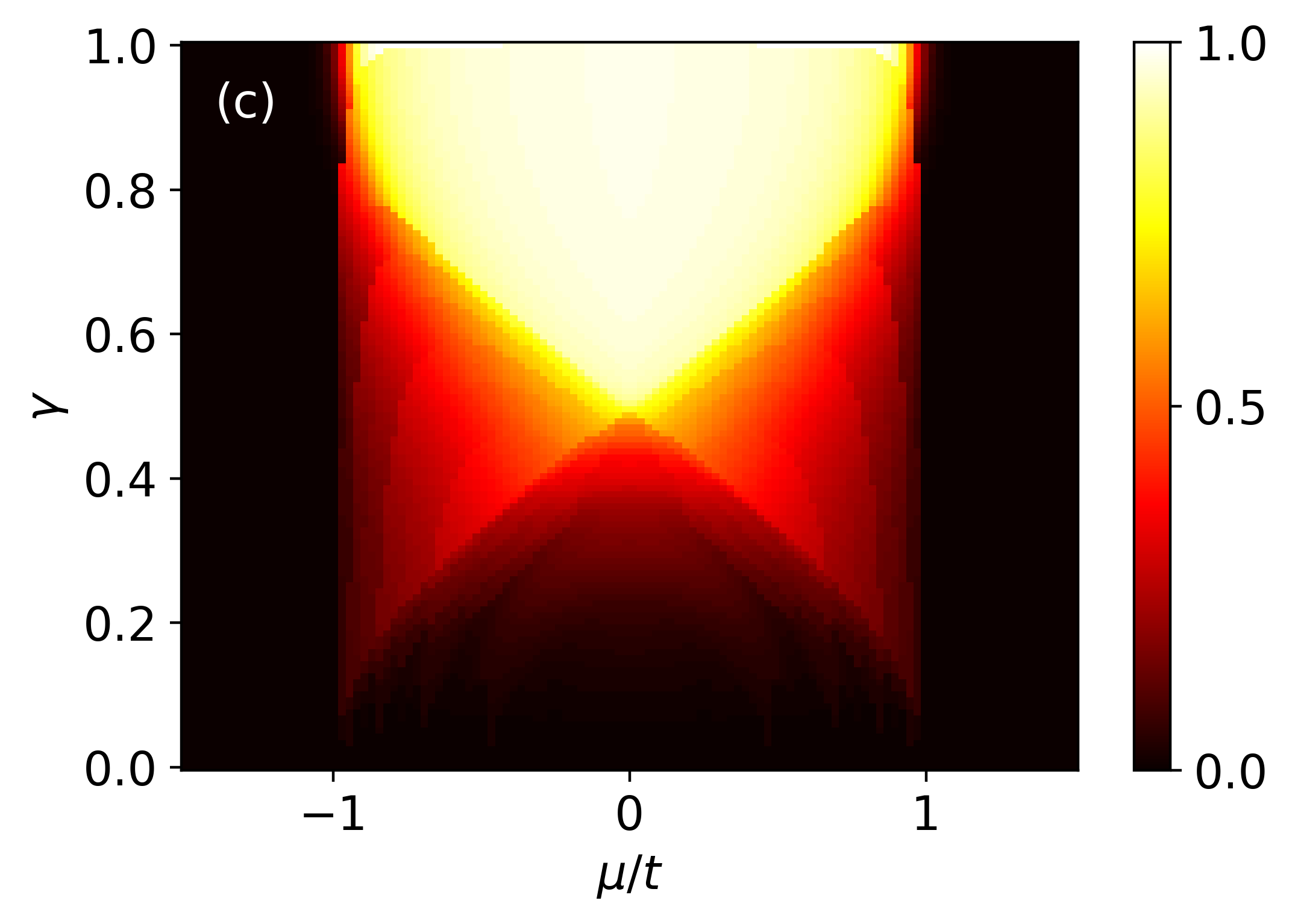}
\includegraphics[width=0.47\linewidth]{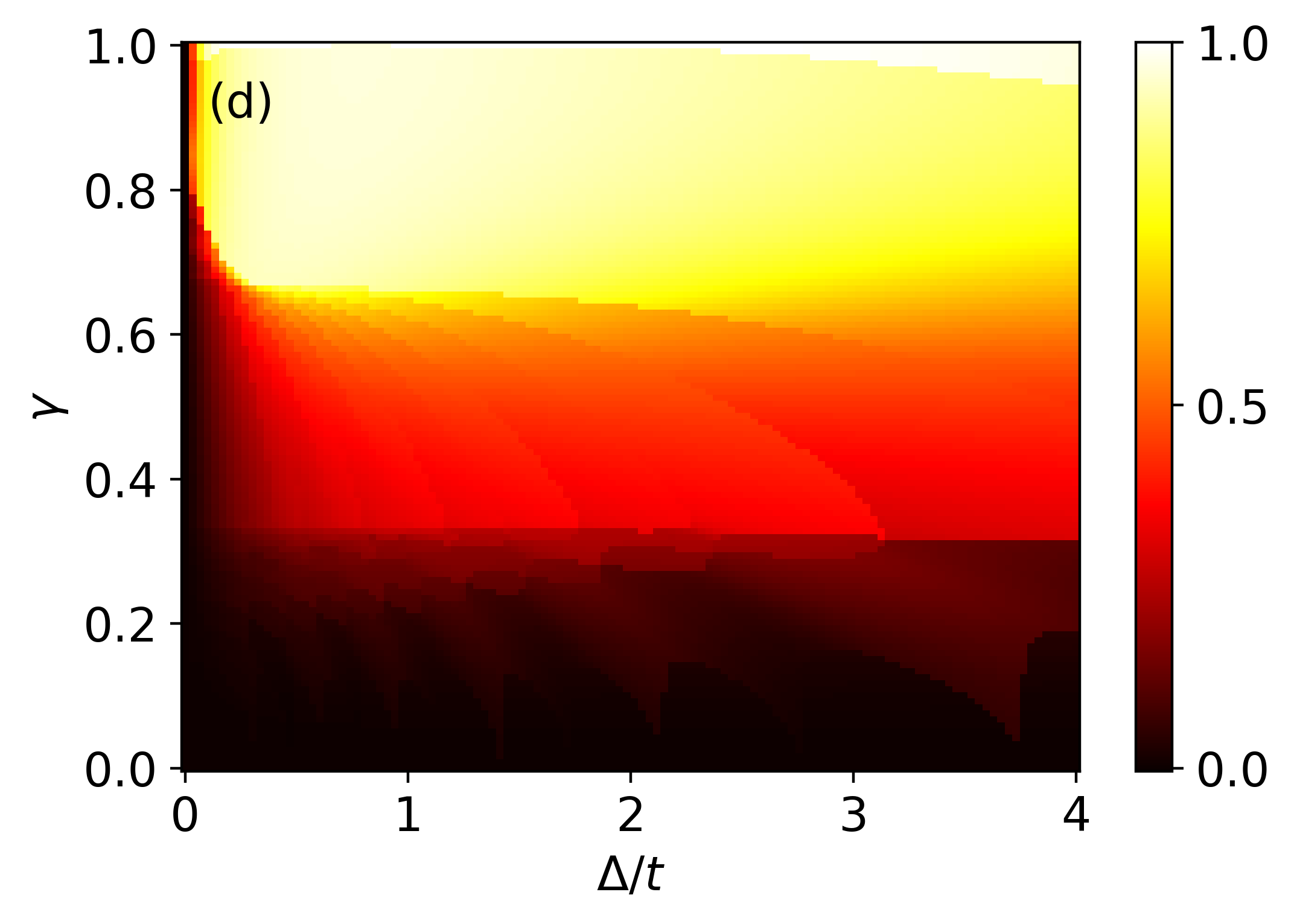}
\caption{Heatmap of winding number in the $(\gamma,\mu/t)\big\vert_{\Delta=1}$ and $(\gamma,\Delta/t)\big\vert_{\mu=1/2}$ planes of phase space for (a), (b) fractional pairing, $\gamma\equiv \gamma_p$ and (c), (d) fractional hopping, $\gamma\equiv \gamma_h$.}
\label{fig:phase}
\end{figure*}

We first concentrate on flat-band Hamiltonians which occur at vanishing chemical potential. We will orient ourselves with respect to the $ \gamma =1$ case where $\mu=0$ and $\Delta\neq 0$ correspond to a topologically non-trivial phase of the Majorana chain. We compare in Fig.\ref{fig:1} the spectral evolution under $\gamma\in (0,1]$ to the familiar parameterization of the ordinary Kitaev chain phase space. For completeness, we provide also the $(\gamma,\mu)$ and $(\gamma,\Delta)$ cuts of the topological phase space probed by $\nu$ in Fig.~\ref{fig:phase}(a,b). For $\gamma>\gamma^*$, the phase diagrams of the ordinary and fractionally twisted Kitaev chain are identical.  


The most notable feature of the spectral evolution as a function of  $ \gamma $ is the transition to a metallic region with rational winding number for $ \gamma < \gamma ^*\equiv 1/2$. Although this metallic state is surprising in Dirac systems with two independent moduli which do not afford any additional tunable parameters that allow the bands to cross generically, this metallic state is in accordance with the von Neumann–Wigner theorem: a Hamiltonian describing dynamics in an $m$ dimensional phase space will have level crossing on an $m-2$ dimensional manifold. In our case, the Hamiltonian has blocks valued in the modified algebra of Eq.~\eqref{eq:comm}, which introduces a third independent parameter, $\gamma$, controlling the central element. Hence, levels can cross on curves (rather than a point) in the moduli space of the Kitaev chain spectra. 


For $ \gamma > \gamma ^*$, the topological phase remains robust, as indicated by the winding number plateau at $\nu=+1$. A stable pair of $E=0$ boundary modes indicates that the bulk-boundary correspondence is intact in this regime. The boundary Majorana modes delocalize as $\gamma\to \gamma^*$ from above as evidenced by the vanishing inverse-participation ratio\begin{footnote}{The inverse participation ratio, $\mathcal I_q:=\sum_x|\psi(x)|^{2q}$, is a measure of the extent of localization of state $\psi$. In the present context, with $q=2$, it is a direct measure of the inverse localization length. A perfectly localized state has IPR 1 while a perfectly delocalized state has IPR 0.}\end{footnote} of the mid-gap mode. This is not, however, accompanied by a vanishing winding number. One normally expects $\nu\to 0^-$ and $\nu\to 1^+$ at a topological phase transition as we see in Fig.~\ref{fig:1}(a,b). Curiously, the winding number interpolates between $+1$ and $0$ in the delocalized regime and equals $1/2$ when the gap closes and the IPR of the zero-mode vanishes. If we can interpret the rational winding number as a form of topological obstruction, the delocalized phase breaks the bulk-boundary correspondence, as the presence of the metallic modes are insensitive to the boundary.


Integer winding numbers are conventionally tied to the existence of distinct static bands. Standard theory suggests that if bands cross or touch, the winding of the ground state is no longer unambiguously connected to some topological invariant of a Hilbert sub-bundle (or an index). In the present case, however, there do exist well defined band edges (see Fig.~\ref{fig:1}(d)), albeit with a continuum of midgap modes. We claim that the rational winding numbers obtained reflect the rational index of $\slashed D_\gamma$ as in the discussion above.

Fig~\ref{fig:phase}(c,d) depicts the phase diagram of the case where $\gamma_h\equiv\gamma$, $\gamma_p=1$ in Eq.~\eqref{eq:kit-domain}. Common to both configurations of the model is a robust regime of integral winding number that transition to domains of rational winding number with the transition happening consistently at $\nu=1/2$. The case where the hopping is fractionalized further indicates that there exists a $\gamma^{**}\sim 1/3$ that separates two regimes of rational winding number with a kink. We defer a detailed analysis of the implications of this feature to later work. We do, however, interpret the emergence of a seemingly critical rational winding number as our first signal that the pseudo-metallic regimes carry some semblance of the topology of the $\gamma=1$ theory.




\subsubsection*{The non-triviality of the pseudo-metallic phase}

\begin{figure}
\includegraphics[width=0.99\linewidth]{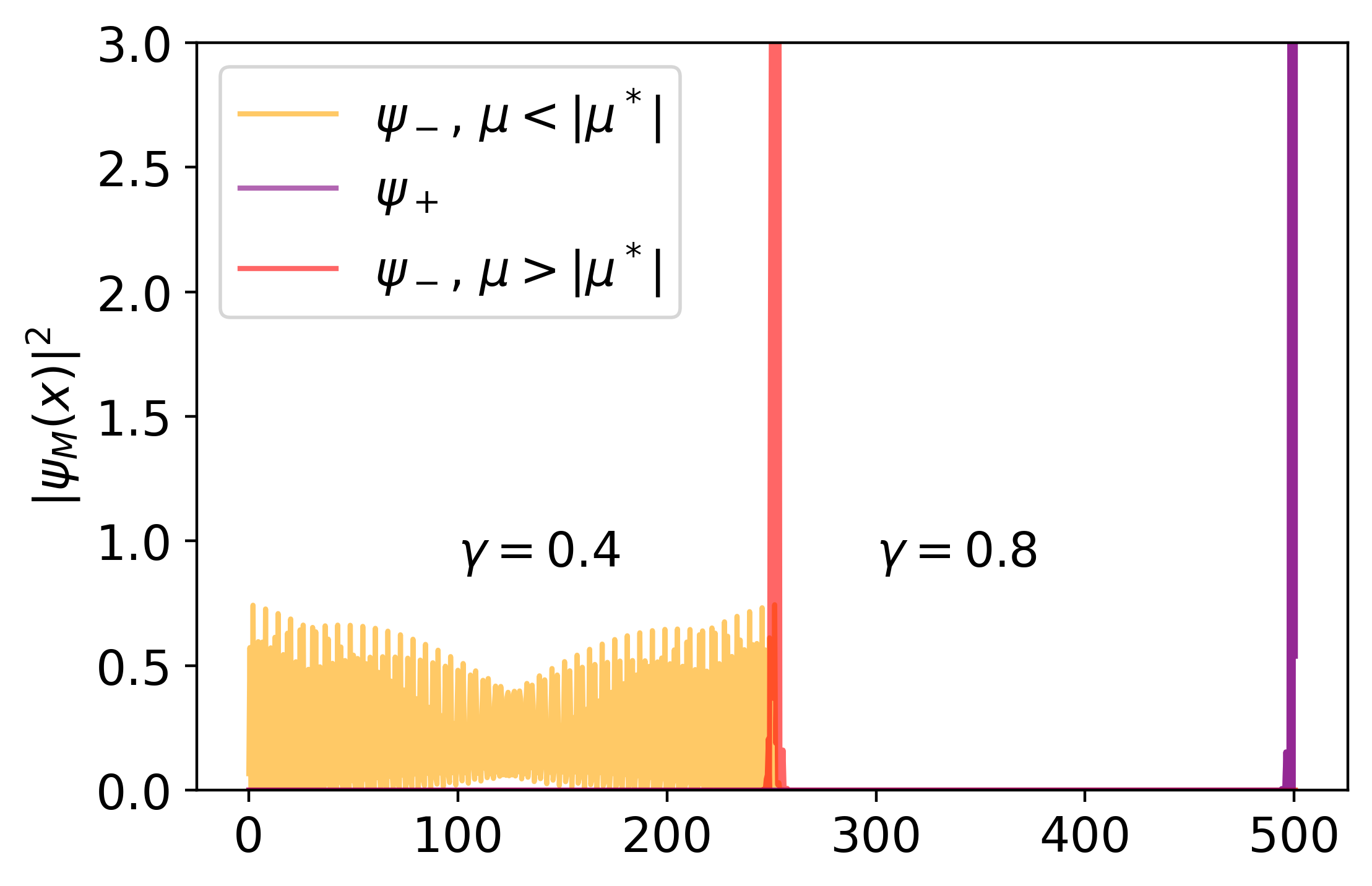}
\caption{$|\psi_M(x)|^2$ corresponding to the two states closest to $E=0$, denoted $\psi_{\pm}$. One of these states is delocalized over the region with $\gamma<\gamma^*$ while the other is localized at the boundary of the region with $\gamma>\gamma^*$. The left domain has a modulated onsite potential that can switch between topological and trivial phases. }
\label{fig:domain}
\end{figure}

Returning to the case of fractional pairing, we further probe the spectrum in the delocalized regime by partitioning the chain into two regions with different values of  $ \gamma $ that straddle $ \gamma ^*$. With open boundary conditions, this corresponds to a Kitaev chain where the right half-chain is in the well understood insulating $\nu=1$ phase and the left half-ring hosts the delocalized mid-gap modes. Tracking the two modes closest to $E=0$ along the chain in Fig.~\ref{fig:domain}, we find boundary localization on the right and delocalization across the left half-chain. When we tune the pseudo-metallic left half-chain to be topologically trivial with $\mu>|\mu^*|$ while maintaining the right half chain at $\nu=1$, we notice that there exists a mid-gap delocalized mode in the spectrum that coalesces into a Majorana mode localized at the domain boundary. 

\begin{figure}
\includegraphics[width=0.95\linewidth]{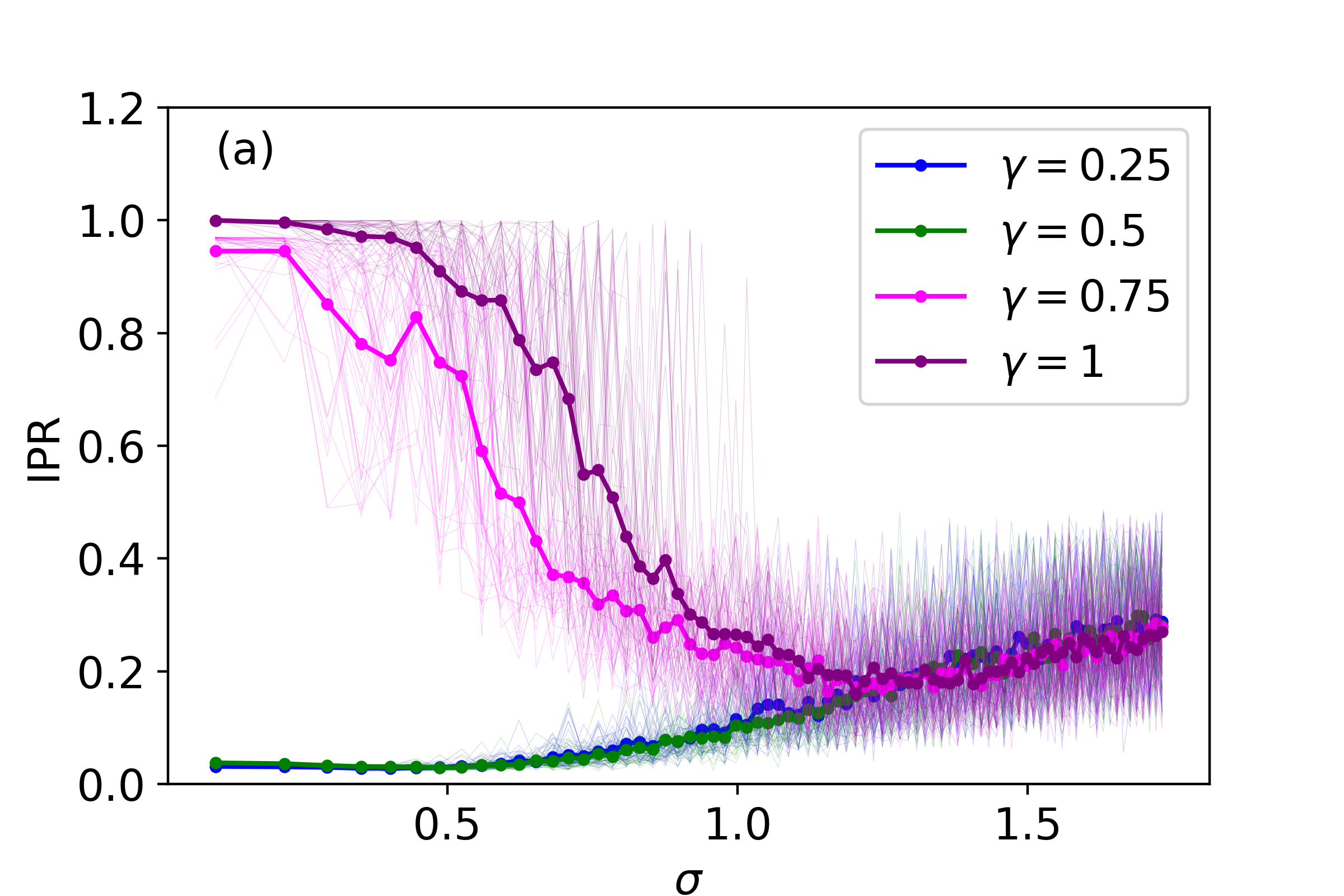}
\includegraphics[width=0.95\linewidth]{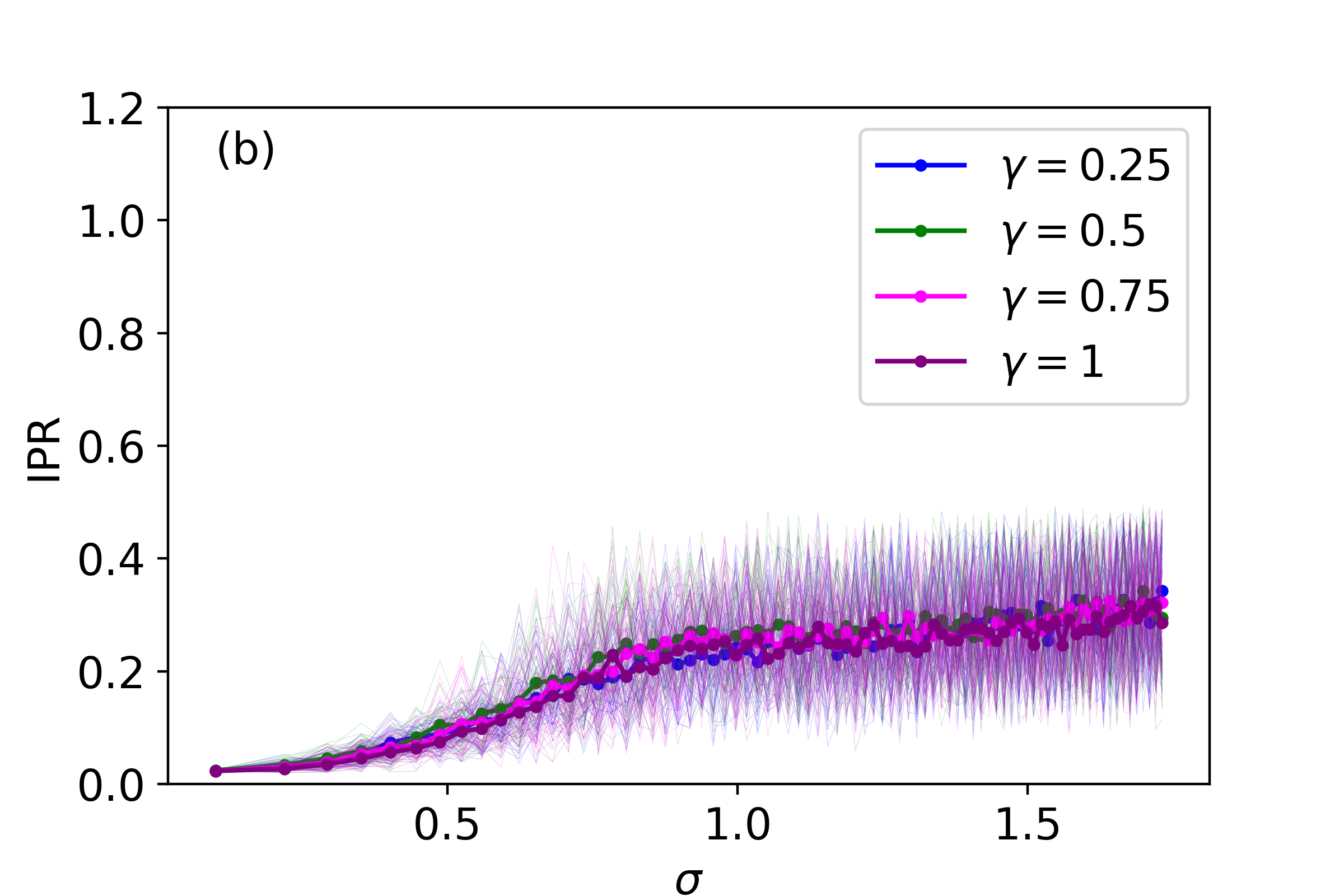}
\caption{Inverse participation ratio for increasing normal disorder variance for characteristic choices of $\gamma$ for (a) the topological phase with $\mu=0$ and (b) the trivial phase with $\mu=2$. The light lines indicate the evolution of ensembles of the disorder realizations and the bold lines indicate ensemble averages.}
\label{fig:ipr}
\end{figure}

This result is necessary but insufficient as an argument for the topological non-triviality of a rational winding number. Next, we introduce onsite disorder in the form of a random Gaussian vector with mean $\mu$ (the chemical potential) and variance $\sigma^2$. That is, we have 
\beq
H_{\text{onsite}}=\tilde \mu \sigma_z;\quad \tilde\mu\sim\mathcal N(\mu,\sigma^2)
\eeq
for each diagonal block of the Hamiltonian. We compute the inverse participation ratio of the midgap state for different values of $\gamma$.  Notably, when $\mu<|\mu^*|$, the delocalized mode that is in correspondence with a Majorana mode in the $\gamma>\gamma^*$ phase initially resists localization. The $\sigma$ for which Anderson localization\cite{loc} obtains is precisely where the localized boundary modes present when $\gamma>\gamma^*$ take the form of generic randomly localized states. When the chemical potential is tuned to the trivial phase where a Majorana modes is absent for $\gamma>\gamma^*$ but  midgap states exist for $\gamma<\gamma^*$, the onset of Anderson physics is immediate at $\sigma>0$ with no sensitivity to the choice of $\gamma$. These facts are illustrated in Fig.~\ref{fig:ipr}. Note that the two choices of mean chemical potential, $\mu=0$ and $\mu=2$ have the same band-gap in the clean limit.

We conclude, therefore, that while the bulk-boundary correspondence is lost for $\gamma<\gamma^*$, there exist a particular pair of extended states that are in correspondence with the topological boundary modes present when $\gamma>\gamma^*$. This is a strong indication that the $\mathbb Q$ valued winding number should be taken seriously as an indicator of non-trivial bulk topology in the pseudo-metallic phase.



\section{Field theoretic perspective}\label{sec:QFT}

Recall here that a Dirac bundle over a Riemannian manifold is a bundle of Clifford modules equipped with a Clifford connection. One can define the Dirac operator, $D$, as a first-order differential operator acting on the space of sections of such bundles. Further, by equipping the Dirac bundle with the compatible action of the Clifford algebra, $\mathfrak {Cl}$, one defines on its sections the operator $\slashed D:= s \cdot D$, $ s \in\mathfrak {Cl}$. It is the index of this operator, which also encodes the constraints of symmetries $\mathcal S,\mathcal T \text{ and }\mathcal C$, that can be related to the topological classification of Dirac-like lattice systems realizing those symmetries. It is therefore to be expected that non-trivially altering the Clifford algebra action (say, by a central extension) will change the index of $\slashed D$. 



Let us begin by defining a theory of Majorana fermions on, for definiteness, $\mathbb C P^1$, equipped with the spin structure, $\Sigma$:
\begin{equation}
    S=i\int_{\mathbb C P^1}\bar\psi \slashed{D}_\Sigma \psi.
\end{equation}
Here, the Dirac operator is chosen to be
\begin{equation}
\slashed{D}_\Sigma = \gamma^1D_1+\gamma^2D_2+\gamma^{3}m,\quad \gamma^{3}=\gamma^1\gamma^2.    
\end{equation}
The free $n$-Majorana path integral is given by the Pffafian,
\begin{equation}
    Z = \text{Pf}(\slashed D_{\mathbb C P^1})^n.
\end{equation}
This theory has two distinct phases, parameterized by mod $2$ $\text{Index}(\slashed D_\Sigma)\in\{1,0\}$. Depending on this (integral) index, 
$$
Z\mapsto (-1)^{\text{Index}(\slashed D_\Sigma)}Z
$$
under $m\mapsto -m$.


The theory under consideration in our note is one where the operator $\slashed{D}_\Sigma$ is replaced by 
\begin{equation}\label{eq:slashDgamma}
    \slashed{D}_{\rho,\Sigma}=\tilde \gamma^1D_1+\tilde \gamma^2D_2+\tilde \gamma^{3}m,
\end{equation}
where $\rho: \mathfrak {Cl}\to PGL(n)$ is a projective representation arising from a central extension of $\mathfrak {Cl}$ and $\tilde \gamma^i=\rho ( \gamma^i)$. In Ref.~\cite{singer} the authors show that when the global manifold $M$ (representing the parameter space of the family of generalized Dirac operators) does not admit a spin$^\mathbb C$ structure but via a torsion class in $H^3(M,\mathbb Z),$ it admits one up to projective representations (or up to central extension). This makes the $\hat A$-genus in the familiar index theorem not an integer but a rational number. 

In turn, under $m\mapsto -m$, the Pfaffian picks up a phase $(-1)^{2s}$, with $s\in\mathbb Q$. Taking $s\equiv p/q$ with $p,q\in\mathbb Z$, one finds that the $q-$Majorana theory with the projective Dirac operator recovers the topological phase structure of the standard $2-$Majorana theory, where the sign change of the Pfaffian is controlled by the number of zero-mode pairs present in the spectrum. This prompts an analogy between the fractional analytic index and a Majorana zero-mode carrying a \emph{fraction} of the topological index.

The intuitive connection to ordinary Majorana physics makes use of the fact that a spin$^\mathbb C$ structure exists whereas one might think the introduction of the projective Dirac operator demands the absence of spin structure. We argue that the salient feature is not that the spin$^\mathbb C$ structure does not exist, but that the (generalized) Dirac operator comes about from a central extension of the Clifford algebra. Hence the rationality of the winding number. We remark that the operator $\slashed{D}_\gamma$ of equation \eqref{eq:slashDgamma} (by virtue of the periodicity in $\gamma \in [0,2]$) defines an operator on $M=\mathbb C \mathbb P^1\times S^1$. This manifold has no torsion classes in $H^3(M,\mathbb Z)$ and indeed does admit a spin$^\mathbb C$ structure, but the arguments of ~\cite{singer} carry through to the case where the Dirac operator does not arise from a spin$^\mathbb C$ but from a projective representation.

\section{Concluding remarks}

The introduction of a twist in the Kitaev Hamiltonian in the form of an operator power of a Pauli matrix centrally extends the Clifford algebra and gives rise to Hamiltonian blocks that are projective representations of the particle-hole symmetry. This turns out to be intimately related to the notion of twisted $K$-theory, which now replaces $K$-theory in the classification of topological materials.  The notable physical consequences of this extension is the appearance of a pseudo-metallic phase within which the winding number is valued in $\mathbb Q$, with the bulk gap closing precisely at $\nu=1/2$. In analogy with the rational analytic index of the projective Dirac operators remaining homotopy invariants~\cite{singer}, we presented numerical evidence in the form of localization resisting metallic modes for the topological non-triviality this new metallic phase.  Finally, we postulated  that along with class BDI topological insulators in dimension 1 (chosen for their simplicity), the entire periodic table is interspersed with non-integral topological indices when the Dirac operator is realized only projectively.  Hence, the scheme posed here, based on Eq. (\ref{newalgebra}) opens up a potentially new route to engineering topological materials.

\textbf{Acknowledgements} 
PWP thanks  DMR-2111379 for partial funding of this project.

\bibliographystyle{apsrev4-1}
%

\end{document}